\documentclass[]{IEEEtran}
\ifCLASSINFOpdf
  \usepackage[pdftex]{graphicx}
\usepackage{amsmath}
\usepackage{amsthm,amssymb}
\usepackage{booktabs}

\ifCLASSOPTIONcompsoc
 \usepackage[caption=false,font=normalsize,labelfont=sf,textfont=sf]{subfig}
\else
 \usepackage[caption=false,font=footnotesize]{subfig}
\fi
\usepackage{algorithmic}
\usepackage{gensymb}
\usepackage[]{algorithm2e}

\usepackage{tabularx}
\usepackage{blindtext}
\usepackage{cite}
\usepackage{booktabs}
\usepackage{pifont}
\usepackage{hyperref}
\usepackage{cleveref}
\usepackage{etoolbox}
\newtoggle{long}
\togglefalse{long}
\newcommand{\iflong}[1]{\iftoggle{long}{#1}{}}
\newcommand{\ifelselong}[2]{\iftoggle{long}{#1}{#2}}

\usepackage{comment}
\newboolean{showcomments}
\setboolean{showcomments}{true}
\newcommand{\zach}[1]{\ifthenelse{\boolean{showcomments}}
{ \textcolor{blue}{(ZL:  #1)}}{}}
\newcommand{\slow}[1]{\ifthenelse{\boolean{showcomments}}
{ \textcolor{red}{(SL:  #1)}}{}}

\newcommand{\beq}{\begin{equation}}
\newcommand{\eeq}{\end{equation}}
\newcommand{\bq}{\begin{eqnarray}}
\newcommand{\eq}{\end{eqnarray}}
\newcommand{\bqn}{\begin{eqnarray*}}
\newcommand{\eqn}{\end{eqnarray*}}
\newcommand{\bee}{\begin{enumerate}}
\newcommand{\eee}{\end{enumerate}}
\newcommand{\bi}{\begin{itemize}}
\newcommand{\ei}{\end{itemize}}

\newcommand{\acndata}{ACN-Data}
\newcommand{\acnsim}{ACN-Sim}
\newcommand{\acnlive}{ACN-Live}
\newcommand{\code}[1]{\texttt{#1}}

\begin{document}
\title{ACN-Sim: An Open-Source Simulator for Data-Driven Electric Vehicle Charging Research}
\author{
\IEEEauthorblockN{Zachary J. Lee\IEEEauthorrefmark{1}, Sunash Sharma\IEEEauthorrefmark{1}, Daniel Johansson\IEEEauthorrefmark{2}, Steven H. Low\IEEEauthorrefmark{1}}\\
\IEEEauthorblockA{\IEEEauthorrefmark{1}Division of Engineering \& Applied Science, Caltech,
Pasadena, CA\\
\IEEEauthorrefmark{2}Faculty of Engineering, Lund University, Lund, Sweden }\\
\IEEEauthorblockA{zlee@caltech.edu}
\vspace{-20pt}
\thanks{This material is based upon work supported by NSF through grants CCF 1637598, ECCS 1619352, and CPS including grant 1739355, as well fellowship support through the NSF Graduate Research Fellowship Program 1745301 and Resnick Sustainability Institute. We would like to thank the team at PowerFlex, especially Cheng Jin, Ted Lee and George Lee, as well as Rand Lee, James Anderson, and Jorn Reniers  for providing data, expertise and ideas to this project.}
}


\maketitle

\begin{abstract}
\acnsim{} is a data-driven, open-source simulation environment designed to accelerate research in the field of smart electric vehicle (EV) charging. It fills the need in this community for a widely available, realistic simulation environment in which researchers can evaluate algorithms and test assumptions. \acnsim{} provides a modular, extensible architecture, which models the complexity of real charging systems, including battery charging behavior and unbalanced three-phase infrastructure. It also integrates with a broader ecosystem of research tools. These include \acndata{}, an open dataset of EV charging sessions, which provides realistic simulation scenarios and \acnlive{}, a framework for field-testing charging algorithms. It also integrates with grid simulators like MATPOWER, PandaPower and OpenDSS, and OpenAI Gym for training reinforcement learning agents.
\end{abstract}

\IEEEpeerreviewmaketitle

\section{Introduction}
With millions of electric vehicles (EVs) expected to enter service in the next decade, generating gigawatt-hours of additional energy demand, engineers must work quickly to develop new algorithms to provide safe and affordable charging at scale. This need has resulted in a large body of research in managed or smart charging algorithms, outlined in \cite{wang_smart_2016, mukherjee_review_2015}.
However, transitioning these algorithms from theory to practice requires dealing with the complexities of practical systems, which are often overlooked in simplified theoretical models. While these simpler models can make analysis tractable, they can also lead to a sizable gap between theoretical results and robust, high-performance implementations of algorithms. Bridging this gap is critical to making an impact in practice, but doing so requires: (1) access to real-world data; (2) detailed simulations driven by realistic models; and (3) the ability to test an algorithm in the field. 

\begin{figure}
    \centering
    \includegraphics[width=.9\columnwidth]{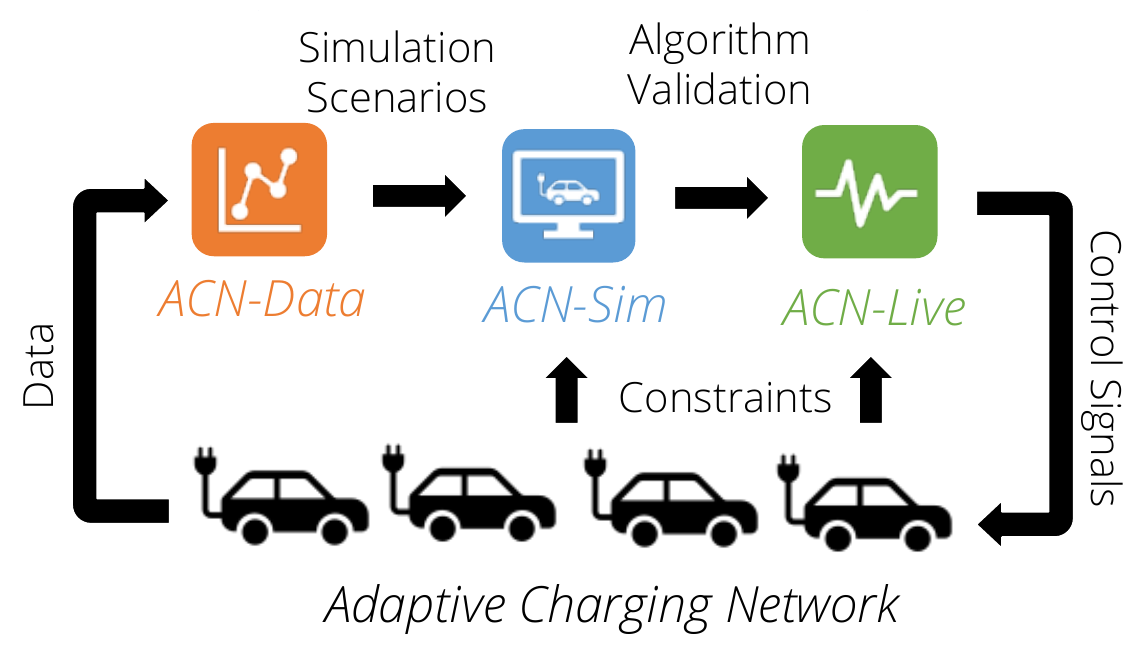}
    \vspace{-0.12in}
    \caption{The ACN Research Portal gives users many of the benefits of an EV charging testbed without needing to build one themselves. It includes data collected from real charging sessions (\acndata{}), a simulator to evaluate new ideas (\acnsim{}), and access to run on real hardware (\acnlive{}).}
    \label{fig:acnportal_overview}
\end{figure}

We began to bridge this gap in 2016 with the development of the Caltech Adaptive Charging Network (ACN), a first-of-its-kind testbed for large-scale, high-density EV charging research \cite{lee_adaptive_2016, lee_large-scale_2018}. This testbed consists of 126 networked and controllable EV charging stations, which allow us to collect data and field test algorithms with real hardware. The ACN Research Portal (ACN-Portal) is designed to give more researchers access to the benefits of this testbed. The portal has three major components: (1) \acndata{}, a dataset of over 66,000 real EV charging sessions from ACNs like the one at Caltech \cite{lee_ACN-Data_2019}; (2) \acnsim{}, an open-source, data-driven simulation environment; and (3) \acnlive{}, a framework for researchers to field test algorithms on the Caltech ACN. The interaction between these tools and the physical ACN infrastructure is summarized in Fig.~\ref{fig:acnportal_overview}. 

In this work, we will focus on \acnsim{}, which is open-source and available at \cite{lee_acnportal_2019}. ACN-Sim has previously been described in \cite{lee2019ACN-Sim_poster, lee2019ACN-Sim}. However, in this work, we extend this description with new features and applications. Our contributions are as follows:
\begin{enumerate}
    \item We describe the architecture and models of \acnsim{} and its integration with MATPOWER, PandaPower, OpenDSS, and OpenAI Gym. 
    \item We describe how algorithms are implemented using \acnsim{}, including a suite of baseline deadline-scheduling algorithms and a new module for model-predictive control, making it easy to implement algorithms like those proposed in \cite{lee_large-scale_2018}.
    \item We demonstrate, using \acnsim{}, that managed EV charging allows charging systems to safely operate with significantly smaller transformers/interconnections and at lower costs than conventional uncontrolled systems.
    \item We compare algorithm performance using \acnsim{}, including the effect of unbalanced three-phase infrastructure, which has not been considered in most smart charging research. 
    \item We evaluate the effect that managed EV charging and onsite solar generation will have on a distribution feeder, using \acnsim{} and its integration with OpenDSS. 
\end{enumerate}

\section{Existing Simulators} \label{sec:existing_work}
Open-source tools and simulators have a long history of supporting smart grid research. MATPOWER\cite{zimmerman_matpower:_2011} makes it easy to solve power flow and optimal power flow problems in MATLAB. It has inspired projects in other languages including PandaPower\cite{thurner_pandapoweropen-source_2018} in Python and PowerModels.jl in Julia\cite{coffrin_powermodels.jl}. Other important simulators include OpenDSS\cite{epri_opendss} and GridLab-D\cite{chassin2014gridlab}, which enable large-scale studies of the distribution system. These tools have demonstrated the importance and impact of open tools within the smart grid community. ACN-Sim integrates with many of these, including MATPOWER, PandaPower, and OpenDSS, to enable studies of the grid impacts of EV charging. 

ACN-Sim is not the first simulator specific to EV charging. V2G-Sim was developed at Lawrence Berkeley National Laboratory and has been used to evaluate the ability of EVs to meet drivers' mobility needs in the context of: level-1 charging \cite{saxena2015charging}; battery degradation \cite{saxena2015quantifying}; and demand response \cite{saxena2015flexibility}. V2G-Sim has also been used to examine grid-level effects of smart charging, such as smoothing the duck curve\cite{coignard2018clean}. EVLib and EVLibSim  were developed at Aristotle University of Thessaloniki to model many types of EV charging, including standard conductive charging, inductive charging, and battery swapping \cite{rigas_evlibsim:_2018}. These simulators address a different problem space from \acnsim{}. While \acnsim{} is designed to evaluate online and closed-loop control strategies, these simulators only allow precomputed schedules or simple controls. \acnsim{} is also unique in modeling unbalanced, behind the meter electrical infrastructure, allowing it to evaluate algorithms which oversubscribe local infrastructure. 

More recently, the Open Platform for Energy Networks (OPEN) from Oxford was released to facilitate simulation and optimization of smart local energy systems, including electric vehicle charging \cite{morstyn_open}.
OPEN supports model predictive control algorithms at the distribution feeder level and unbalanced three-phase infrastructure. It also allows for control of other distributed energy resources such as stationary storage and building loads. However, it has not been used to consider the electrical infrastructure behind the meter.

Overall, \acnsim{}'s realistic models and data taken from real charging systems, along with its simple interfaces for defining new control algorithms and suite of baseline algorithms, set it apart from existing simulators, making it a useful addition to the suite of tools available to researchers.

\begin{figure}
    \centering
\includegraphics[width=\columnwidth]{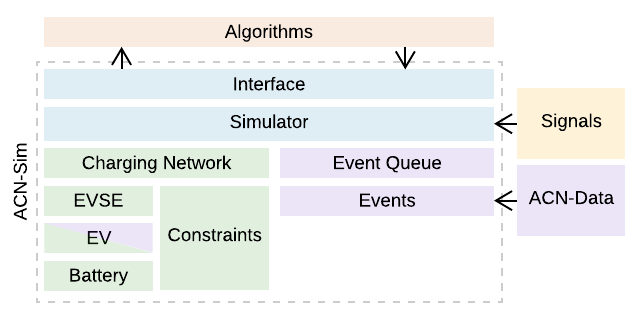}
    \vspace{-0.3in}
    \caption{Architecture of \acnsim{} along with related sub-modules Signals, Algorithms, and ACN-Data. Note that \code{EV} models both the physical vehicle and session information, such as departure time and energy requested.
    }
    \label{fig:acnsim_structure}
\end{figure}

\section{Simulator Architecture and Models}\label{sec:architecture}
\acnsim{} utilizes a modular, object-oriented architecture which is shown in Fig.~\ref{fig:acnsim_structure}. This design models physical systems as closely as possible and makes it easier to extend the simulator for new use cases. Each of the boxes in Fig.~\ref{fig:acnsim_structure} refers to a base class that can be extended to model new behavior or add functionality. While \acnsim{} includes several models of each component, users are free to customize the simulator to meet their needs. We encourage researchers to contribute extensions back to the project so that others can utilize them.

\subsection{Simulator} \label{sec:simulator}

\begin{figure}
    \centering
    \includegraphics[width=0.75\columnwidth]{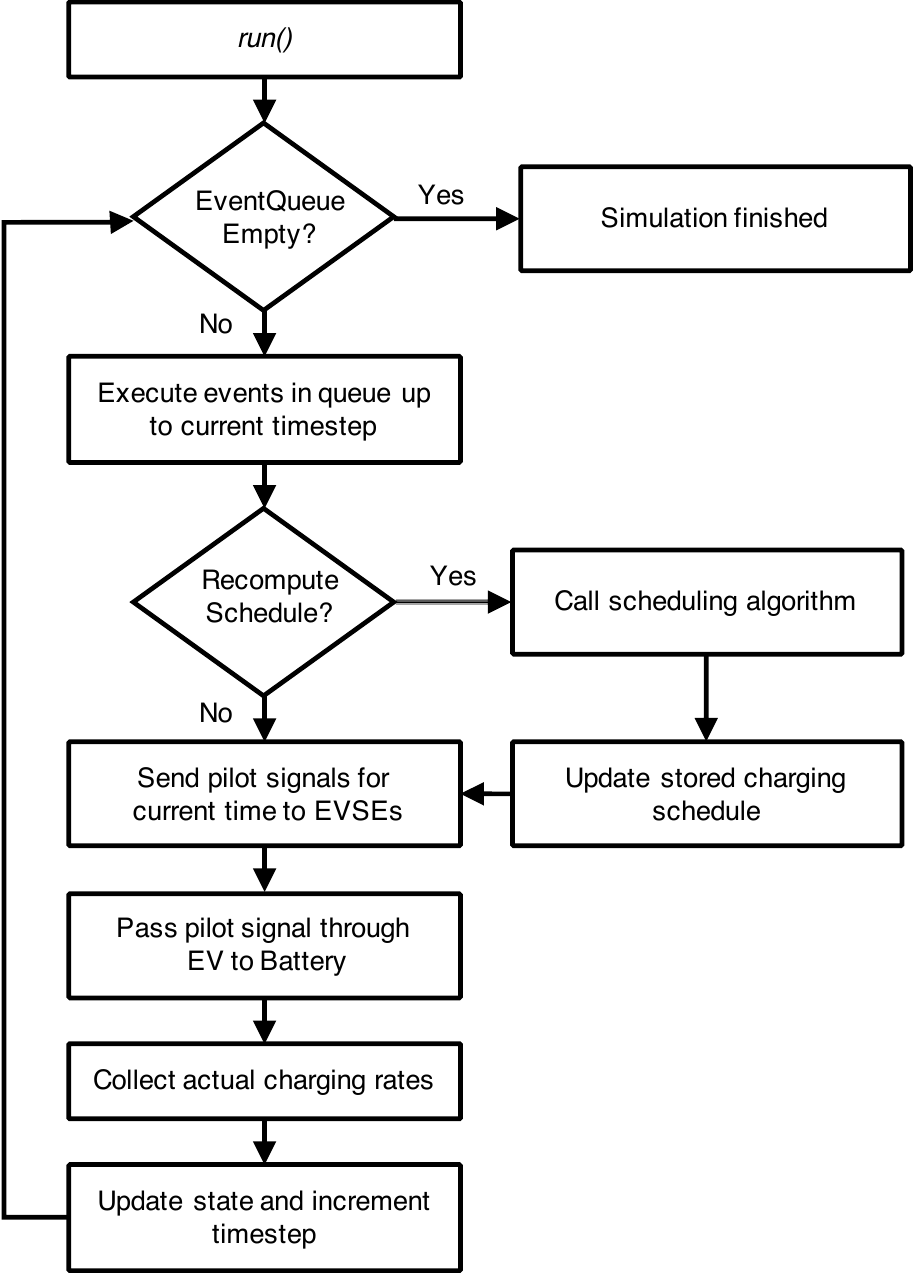}
    \caption{Flow chart describing the simulator's \code{run()} function. Each timestep consists of a single iteration of this loop. The simulation ends when the last event from the \code{EventQueue} is executed at which time the user can analyze the results of the simulation.}
    \label{fig:acnsim_flow}
\end{figure}

A \code{Simulator} object forms the base of any \acnsim{} simulation. This \code{Simulator} holds models of the hardware components in the simulated environment and a queue of events that define when actions occur in the system. ACN-Sim is based on a discrete-time, event-based simulation model. Figure~\ref{fig:acnsim_flow} describes its operation. During a simulation, the \code{Simulator} stores relevant data, such as the event history, EV history, and time series for the pilot signal and charging current for each EVSE, for later analysis.

\subsection{Charging Network}

\subsubsection{Electrical Infrastructure}
\acnsim{} uses the \code{ChargingNetwork} class to model the electrical infrastructure of the charging system, including EVSEs\footnote{EVSE stands for Electric Vehicle Supply Equipment, they are more commonly known as charging stations or charging ports.}, transformers, switch panels, and cables. Each \code{ChargingNetwork} instance contains a set of \code{EVSE} objects, as well as a set of constraints.

We model constraints by limiting the current through each bottleneck component in the network. Because charging systems are radial networks and electrical codes specify ampacity limits that keep voltages within specifications, it is sufficient to model only constraints on current magnitudes.
Using Kirchhoff's Current Law, we can express these constraints by

\begin{equation*}
    |I_j(t)| = \left| \sum_{i=1}^N A_{ij} r_i(t)e^{j\phi_{i}} \right| \leq R_j, \ \  \forall t \in \mathcal{T}
\end{equation*}

\noindent where $I_j(t)$ is the current through the bottleneck, $R_j$ is the limit on the current magnitude, $N$ is the total number of \code{EVSE}s in the system, $r_i(t)$ is the charging current of \code{EVSE} $i$ at time $t$, and $\mathcal{T}$ is the set of all timesteps in the simulation. The parameter $\phi_i$ is the phase angle of the current phasor, which can be calculated based on how EVSE $i$ is connected in the network. For simplicity, we assume $\phi_i$ is fixed, and voltages in the network are nominal. $A_{ij}$ can be found via circuit analysis, as shown in \cite{lee_large-scale_2018} for a subset of the Caltech ACN. 

\iflong{To incorporate these constraints, algorithms can either parse the constraints and include them directly in the algorithm, as is done in model predictive control, or use the built-in \code{is\_feasible()} method, which returns if the proposed charging rates are allowable under the given network model.}

\subsubsection{Stochastic Space Assignment}
\code{ChargingNetwork} assumes that each EV is preassigned to a specific EVSE, and no two EVs are ever assigned to the same EVSE at the same time. This holds when applying a workload from \acndata{} to its corresponding network model. However, in some cases, such as when generating events from a statistical model or applying a real workload to a new network configuration, it can be helpful to allow for non-deterministic space assignments. \acnsim{} accomplishes this through the \code{StochasticNetwork} class (which is a subclass of \code{ChargingNetwork}). Using this network model, EVs are assigned to a random open EVSE when they arrive instead of using a predefined \code{station\_id} for assignment. Since it is possible for no EVSEs to be available when a new EV arrives, \code{StochasticNetwork} also includes a waiting queue for EVs which arrive while all EVSEs are in use. When an EV leaves the system, the first EV in the queue takes its place. By default, we assume that the presence of EVs in the waiting queue does not affect drivers' departure time. However, with the \code{early\_departure} option, drivers swap places with the first EV in the queue as soon as they finish charging. This is a common practice in many offices that have more EV drivers than EVSEs.

\subsubsection{Included Site Models}
While users are free to develop their own charging networks, \acnsim{} includes functions to generate network models that match the physical infrastructure of the three sites currently included in \acndata{} (Caltech, JPL, and Office001). In addition, the \code{auto\_acn} function allows users to quickly build simple single-phase and three-phase networks by providing just a list of station ids and a transformer capacity. In these \code{auto\_acn} networks, it is assumed that the transformer is the only source of constraints. All of these functions work with both \code{ChargingNetwork} and \code{StochasticNetwork}, which can be set as a parameter. 

\subsection{EVSE}
EVSEs, short for Electric Vehicle Supply Equipment, are the devices EVs plug into to charge.
The EVSE communicates a pilot signal to the EV's on-board charger, which is an upper limit on current the EV is allowed to draw from the EVSE. The granularity of this pilot is dependent on the particular EVSE. Some EVSEs provide continuous control, while others offer only a discrete number of set-points. In addition, according to the J1772 standard, no pilot signals are allowed between 0 to 6 A \cite{sae_sae_2017}.  In most current research, the additional constraints imposed by EVSEs without continuous control are neglected \cite{wang_smart_2016}. However, including these constraints is important for practical algorithms and is non-trivial.

\acnsim{} provides three EVSE models which cover most ideal and practical level-2 EVSEs:
\begin{itemize}
    \item \code{EVSE} allows any pilot signal between an upper and lower bound. By default, \code{EVSE} allows any non-negative charging rate.
    \item \code{DeadbandEVSE} also allows continuous pilots but excludes 0 - 6 A as required by the J1772 standard.
    \item \code{FiniteRatesEVSE} only allows pilot signals within a finite set, accurately modeling most commercial EVSEs. 
    For example, many of the EVSEs used in the Caltech ACN allow \{6, 7, ..., 31, 32\} or \{8, 16, 24, 32\} amps. 
\end{itemize}

\iflong{
Within \acnsim{}, \code{EVSE} is also the interface between the charging network and an \code{EV}. When an \code{EV} plugs into the system, a reference to that \code{EV} is added to the corresponding \code{EVSE}. When it is time to update the pilot to an \code{EV}, the \code{Simulator} first passes the pilot to the \code{EVSE}, which in turn passes it on the \code{EV} and eventually the \code{Battery}. This mimics the flow of information in a real charging system. 
Similarly, when an \code{EV} leaves the system, the reference to that \code{EV} is removed from the \code{EVSE}.}

\subsection{EV}

The \code{EV} object contains relevant information for a single charging session such as arrival time, departure time, estimated departure time and requested energy. The estimated departure time may differ from the actual departure time. Likewise, it may be infeasible to deliver the requested energy in the allotted time due to maximum charging rate restrictions, system congestion, or insufficient battery capacity. By allowing this, \acnsim{} models the case where user inputs or predictions are inaccurate, which is common in practice \cite{lee_ACN-Data_2019}.

\subsection{Battery}

Most research in EV charging utilizes an ideal battery model, where EVs are assumed to follow the given pilot signal exactly.
However, in practice, we see that the charging rate of an EV is often strictly lower than the pilot signal and decays as the battery approaches 100\% state-of-charge \cite{wang_smart_2016, lee_large-scale_2018}. This can significantly increase the total time required to charge the battery and results in under-utilization of infrastructure capacity.

\acnsim{} jointly models the vehicle's battery and battery management system. The battery's actual charging rate depends on the pilot signal as well as the vehicle's on-board charger, its state-of-charge, and other environmental factors. \acnsim{} currently includes two battery models. 

The \code{Battery} class is an idealized model and serves as the base for all other battery models. The actual charging rate of the battery, $\hat{r}(t)$, in this idealized model is described by
\begin{equation*}
    \hat{r}(t) := \min\{r(t),\  \bar{r},\  \hat{e}(t)\}
\end{equation*}

\noindent where $r(t)$ is the pilot signal passed to the battery, $\bar{r}$ is the maximum charging rate of the on-board charger and $\hat{e}(t)$ is the difference between the capacity of the battery and the energy stored in it at time $t$ in the units of A$\cdot$periods. We do not consider discharging batteries, so all rates are positive.
    
\code{Linear2StageBattery} is an extension of \code{Battery} that approximates the roughly piecewise linear charging process used for lithium-ion batteries. The first stage, referred to as \emph{bulk charging}, typically lasts from 0\% to between 70 to 90\% state-of-charge. During this stage, the current draw, neglecting changes in the pilot, is nearly constant. In the second stage, called \emph{absorption}, the voltage of the battery is held constant while the charging current decreases roughly linearly. The actual charging rate of the \texttt{Linear2StageBattery} is given by

\begin{equation*}
\hat{r}(t) := \hfill
\begin{cases}
\min\{r(t),\  \bar{r},\  \hat{e}(t)\}& \text{if } \texttt{SoC} \leq \texttt{th} \\
\min \left\{ \left (1 - \texttt{SoC} \right)\frac{\bar{r}}{1 - \texttt{th}} \text{, } r(t) \right\}  & \text{otherwise}
\end{cases}    
\end{equation*}

\noindent where  \texttt{SoC} is the state-of-charge of the battery and \texttt{th} marks the transition from the \emph{bulk} stage to the \emph{absorption} stage of the charging process.
Figure~\ref{fig:battery_model} shows how these two models compare for two charging profiles taken from \acndata{}.
\ifelselong{

In general, we find that while the piecewise linear model is a good approximation, it does not capture some battery dynamics that we observe in some charging sessions (as in the right panel of Fig.~\ref{fig:battery_model}). Because of this, we plan to implement additional, higher fidelity battery models in future releases of \acnsim{}.}{
In general, we find that the piecewise linear model is a good, though imperfect, approximation.
}
 
\begin{figure}
    \centering
    \includegraphics[width=\columnwidth]{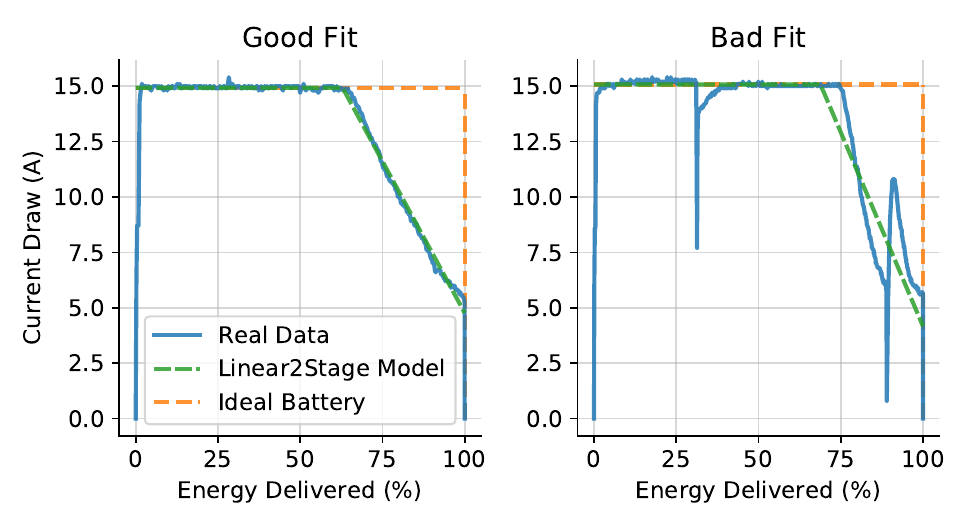}
    \vspace{-0.25in}
    \caption{Comparison of \code{Linear2Stage} and idealized \code{Battery} models with a real charging curve collected from two distinct users of the Caltech ACN when the pilot signal is not binding. We can see that the \code{Linear2Stage} model with appropriate parameters matches the battery behavior well in the first case, but in the second case, there are dynamics in the joint battery/battery manager system that the \code{Linear2Stage} model does not capture; namely the double-tail behavior (which is recurring for this user).}
    \label{fig:battery_model}
\end{figure}

\vspace{-0.025in}
\subsection{Event Queue / Events}
\vspace{-0.025in}
\acnsim{} uses events to describe actions in the simulation. There are two types of events currently supported:
\begin{itemize}
    \item \code{PluginEvent} signals when a new EV arrives at the system. A \code{PluginEvent} also contains a reference to the \code{EV} object which represents the new session.
    \item \code{UnplugEvent} signals when an EV leaves the system at the end of its charging session. 
\end{itemize}
Each event has a timestamp describing when the event should occur. Events are stored in a queue sorted by their timestamp. Since multiple events could occur at the same timestep, we further sort by event type, first executing \code{UnplugEvents}, then \code{PluginEvents}. At each timestep, the \code{Simulator} executes all events left in the queue with timestamps on or before the current timestep.  After any event, the scheduling algorithm is called to adapt to the new system state. Users are free to create new events by extending the \code{Event} class.

To generate events, users can either get real event sequences from \acndata{}, generate event sequences from statistical models, or manually create events to investigate edge cases. To make accessing \acndata{} simpler for users, \acnsim{} provides direct integration with the \acndata{} API. It also provides utilities for learning statistical models such as Gaussian Mixture Models, directly from data using tools from SciKit Learn as described in \cite{lee_ACN-Data_2019}.

\subsection{Signals}
The signals sub-module allows ACN-Sim to integrate with external signal sources, which can be an important part of EV charging systems such as: 1) utility tariffs, 2) solar generation curves, 3) external loads. 

\ifelselong{

\subsubsection{Utility Tariffs}
Operating costs are an important concern for EV charging facilities. To make calculating these costs as simple as possible, ACN-Sim allows users to select from a collection of real tariff structures or import their own in JSON format. Seasonal time-of-use rates and demand charge are both supported. This functionality allows users to investigate cost minimization strategies and accurately estimate the operating costs of charging system designs under different tariff structures.

\subsubsection{Solar Generation}
As many sites with EV charging also have onsite solar generation, studying the behavior of an EV charging facility that takes solar generation into account is an important use case. \acnsim{} allows users to input a solar generation signal as a CSV file to the Simulator. Solar data can be user-generated, downloaded from an external source, or generated by an external solar generation simulator such as NREL's system advisor model (SAM)\cite{blair2018system}. Such functionality allows users to study the effects of onsite solar on cost, energy demand, grid loading, and other metrics associated with large-scale EV charging. 

\subsubsection{External Load}
EV charging facilities often share a meter with other loads, such as the buildings on a university or corporate campus. To reduce demand charge and stress on the grid, it can be advantageous to consider these other loads when scheduling EV charging. To facilitate the study of algorithms that do this, \acnsim{} allows users to input an external load profile as a CSV file. External load data can be user-generated or downloaded from an external source. 
}{
To support utility tariffs, \acnsim{} includes the \code{TimeOfUseTariff} class, which supports time-varying and seasonal tariff schedules with or without demand charges. To make integration easier for users, \acnsim{} includes several utility tariff schedules. Users can define new schedules in a simple JSON format. In Section~\ref{sec:system_planning} we use these tariff schedules to calculate operating costs, and in Section~\ref{sec:oversubscribed_infrastructure} we provide the tariff as an input to the minimum cost objective for the MPC algorithm. 

External loads, generation profiles, pollution indexes, and other signals can be loaded into the simulator using the \code{signals} dictionary within the \code{Simulator} constructor or passed directly to the control algorithm. Section~\ref{sec:grid} provides an example of an experiment using external loads and generation.
}

\subsection{Co-simulation with Grid Simulators}
\acnsim{} also provides co-simulation with popular grid-level simulators, including MATPOWER, PandaPower, and OpenDSS. This allows researchers to investigate Vehicle-Grid Integration (VGI) problems such as algorithms to alleviate voltage and overload issues in the local distribution system or aggregation approaches to bid into markets. In the current version, simulations are run sequentially, with the output of the \acnsim{} experiment serving as an input to the grid simulator. In future releases, we plan to support feedback from the grid simulation into \acnsim{}. 

\subsection{OpenAI Gym Integration}
Reinforcement learning (RL) has been applied to many problems in scheduling and resource allocation \cite{kaelbling_littman_moore_1996}. To enable researchers to study RL algorithms' performance on scheduled EV charging, \acnsim{} provides integration with OpenAI Gym \cite{brockman2016openai}, a framework for testing RL algorithms on different environments. \acnsim{} offers customizable OpenAI Gym environments that implement the same interface as built-in Gym environments \cite{sharma_gym_acnportal_2020}, facilitating both direct application of baseline RL algorithms and development of new algorithms for the problem of smart EV charging. As with other parts of \acnsim{}, users may extend the base \acnsim{} environment to study different action, observation, or reward structures.

\section{Charging Algorithms} \label{sec:algorithms}

\subsection{Interface}

To make algorithm implementations more flexible, we introduce an interface that abstracts away the underlying infrastructure, whether that be simulated or real, allowing us to use the same algorithm implementation with both \acnsim{} and \acnlive{}. This means that algorithms can be thoroughly tested with \acnsim{} before they are used on physical hardware. \iflong{It also means algorithms developed to work with \acnsim{} can work with other platforms simply by extending the \code{Interface} class.}

\subsection{Defining an algorithm}
To define an algorithm in \acnsim{} users only need to extend the \texttt{BaseAlgorithm} class and define the \texttt{schedule()} function. This function takes in a list of active sessions, meaning that the EV is plugged in and its energy demand has not been met, and returns a charging schedule for each. 
\iflong{This schedule is a dictionary that maps station ids to a list of charging rates in amps. }
Each entry in the schedule is valid for one timestep beginning at the current time. Algorithms have access to additional information about the simulation through the \code{Interface} class, such as the current timestep, infrastructure constraints, and allowable pilot signals for each EVSE.

\subsection{Included algorithms}\label{sec:included_alg}

\acnsim{} is packaged with many common online scheduling algorithms that can be used as benchmarks.

\begin{itemize}
    \item \textbf{Uncontrolled Charging}: Most charging systems today do not manage charging. With Uncontrolled Charging, each EV charges at its maximum allowable rate. This algorithm does not factor in infrastructure constraints. 
    \item \textbf{Round Robin}: Round Robin (RR) is a simple algorithm that attempts to equally share charging capacity among all active EVs. It creates a queue of all active EVs. For each EV in the queue, it checks if it is feasible to increment its charging rate by one unit. If it is, it increments the rate and replaces the EV at the end of the queue. If it is not, the charging rate of the EV is fixed, and the algorithm does not return the EV to the queue. This continues until the queue of EVs is empty.
    \iflong{In this context, a feasible charging rate is one which does not cause an infrastructure constraint to be violated and is less than the maximum charging rate of the EVSE.}
    \item \textbf{Sorting Based Algorithms}: Sorting based algorithms are commonly used in other deadline scheduling tasks such as job scheduling in servers due to their simplicity \cite{stankovic2012deadline}. \acnsim{} includes several of these algorithms, including First-Come First-Served (FCFS), Last-Come First-Served (LCFS), Earliest-Deadline First (EDF), Longest Remaining Processing Time (LRPT), and Least-Laxity First (LLF). These algorithms work by first sorting the active EVs by the given metric, then processing them in order. Each EV is assigned its maximum feasible charging rate, given that the assignments to all previous EVs are fixed. This process continues until all EVs have been processed.

    \item \textbf{Model Predictive Control}: Many approaches to the EV scheduling problem rely on model predictive control (MPC). The adacharge package, available at \cite{lee_adacharge_2019}, is based on CVXPY \cite{diamond2016cvxpy, agrawal2018rewriting}, and makes it easy to use these algorithms with \acnsim{}. With this library, users can easily choose from existing objective functions and constraints or create their own. The general framework for these MPC algorithms is outlined in \cite{lee_large-scale_2018}. 
\end{itemize}

\section{Use Cases}
\acnsim{} has been used to explore many research questions. In this section, we provide examples, including evaluating (1) possible infrastructure solutions, (2) the effect of unbalance on oversubscribing infrastructure, (3) time-series of EV charging profiles, and (4) the effect of large-scale EV charging on a distribution feeder. In addition, \acnsim{} has been used to design dynamic pricing schemes and cost-optimal scheduling \cite{lee_pricing_2020}, train reinforcement learning agents for EV charging systems \cite{zishan_adaptive_control}, and examine the effect of non-ideal batteries and EVSE pilot quantization on model predictive control and baseline algorithms \cite{lee_adaptive_charging_networks}.
The code for all case studies presented here is available at \cite{lee_acnsim_demo_2019}. 

\subsection{System Planning}\label{sec:system_planning}
In this section, we demonstrate how the simulator can be used to aid in system planning and design. We consider a site host who would like to install an EV charging solution at an office building. The host estimates that the system will charge approximately 100 EVs per day. 
\ifelselong{
There are several ways to meet this demand: 1) install 102 (34 per phase) level-1 EVSEs with a maximum charging power of 1.9 kW along with a 200 kW transformer, 2) install 102 level-2 EVSE with a maximum charging power of 6.7 kW along with a 685 kW transformer, 3) install 30 level-2 EVSEs along with the 200 kW transformer, 4) install 102 level-2 EVSEs with the 200 kW transformer and use smart charging algorithms to avoid overloading the transformer. Each of these options has its own trade-offs.}{
There are several ways to meet this demand, which are described in Table~\ref{tab:infrastructure_tradeoffs}.
}
Each of these options has its own trade-offs.

We can use ACN-Sim to guide this site host. We assume that the office will have a usage pattern similar to that of JPL. As such, we train a Gaussian Mixture Model based on the data collected from weekday usage at JPL, as described in \cite{lee_ACN-Data_2019}. We assume the site will not allow usage on weekends. We then use ACN-Sim's \code{GaussianMixtureEvents} tool to create a queue of events from this generative model, assuming 100 arrivals on weekdays and 0 on weekends. We also create models of the charging networks described in each proposal. Since EVs are generated, we use the \code{StocasticNetwork}, which randomly assigns EVs to EVSEs when they arrive. For proposals 1, 2, and 3, we use the built-in \code{Uncontrolled} charging algorithm. For proposal 4, we consider an MPC based algorithm for cost minimization. We evaluate the scenarios on four criteria: 1) transformer capacity required, 2) percentage of total energy requested that was delivered, 3) number of times drivers need to swap spaces to allow others to charge after they finish, and 4) the operating cost of the system based on the summer rates from the \code{sce\_tou\_ev\_4\_march\_2019} tariff schedule included in \acnsim{}. We repeat these experiments for ten months of generated data, with mean results shown in Table~\ref{tab:infrastructure_tradeoffs}. Note that in each case, the standard deviation between months was less than $3.5\%$ for each metric. 

\begin{table}[]
\caption{Infrastructure Solution Evaluation (100 EV / Day)}\label{tab:infrastructure_tradeoffs}
\resizebox{\columnwidth}{!}{%
\begin{tabular}{@{}cccc|ccc@{}}
\toprule
\begin{tabular}[c]{@{}c@{}}EVSEs\\ (\#)\end{tabular} & \begin{tabular}[c]{@{}c@{}}EVSE \\ Type\end{tabular} & Alg & \begin{tabular}[c]{@{}c@{}}Transformer \\ Capacity\end{tabular} & \begin{tabular}[c]{@{}c@{}}Swaps\\ (\#/month)\end{tabular} & \begin{tabular}[c]{@{}c@{}}Demand \\ Met\end{tabular} & \begin{tabular}[c]{@{}c@{}}Cost\\ (\$/kWh)\end{tabular} \\ \midrule
102 & Level 1 & Unctrl & 200 & 0 & 75.4\% & 0.278 \\
102 & Level 2 & Unctrl & 685 & 0 & 99.9\% & 0.351 \\
30 & Level 2 & Unctrl & 200 & 1,103.5 & 99.6\% & 0.256 \\
102 & Level 2 & MPC & 200 & 0 & 99.8\% & 0.234 \\ \bottomrule
\end{tabular}
}
\\
\\
\caption{Infrastructure Solution Evaluation (200 EV / Day)}\label{tab:infrastructure_tradeoffs_200}
\resizebox{\columnwidth}{!}{%
\begin{tabular}{@{}cccc|ccc@{}}
\toprule
\begin{tabular}[c]{@{}c@{}}EVSEs\\ (\#)\end{tabular} & \begin{tabular}[c]{@{}c@{}}EVSE \\ Type\end{tabular} & Alg & \begin{tabular}[c]{@{}c@{}}Transformer \\ Capacity\end{tabular} & \begin{tabular}[c]{@{}c@{}}Swaps\\ (\#/month)\end{tabular} & \begin{tabular}[c]{@{}c@{}}Demand \\ Met\end{tabular} & \begin{tabular}[c]{@{}c@{}}Cost\\ (\$/kWh)\end{tabular} \\ \midrule
102 & Level 1 & Unctrl & 200 & 1,174.5 & 73.2\% & 0.244 \\
102 & Level 2 & Unctrl & 680 & 1,081.5 & 99.8\% & 0.327 \\
30 & Level 2 & Unctrl & 200 & 2,973.9 & 91.6\% & 0.233 \\
102 & Level 2 & MPC & 200 & 1,441.9 & 87.1\% & 0.223 \\
201 & Level 2 & MPC & 200 & 0 & 98.4\% & 0.227\\
\bottomrule
\end{tabular}
}

\end{table}

From Table~\ref{tab:infrastructure_tradeoffs}, we can see that while installing 100 level-1 EVSEs might be the simplest solution, these slow chargers are only able to 75.6\% of demand because they cannot support users with large energy needs and short deadlines. However, the alternative of installing a 685 kW transformer and associated service upgrade would be cost-prohibitive for most sites, and installing only 30 level-2 EVSEs requires over 1,100 swaps per month, leading to lost productivity and poor user experience. In this case, the smart charging solution with model predictive control has clear advantages in both capital cost (only requiring a 200 kW transformer), user satisfaction (no swaps are necessary while nearly all user demands are met), and operating costs (having the lowest cost per kWh). This illustrates the real-world need for smart charging systems and associated algorithms.

The benefit of smart charging approaches is amplified as EV adoption grows, and charging infrastructure must scale accordingly. In this scenario, we consider how the system will scale to 200 charging sessions per day. The results are shown in Table~\ref{tab:infrastructure_tradeoffs_200}. Intuitively the systems designed for 100 EVs per day require far more swaps with increased demand, and similarly, the percent of demand met decreases. This is also true for the smart charging (MPC) case. However, while scaling the number of EVSEs in traditional uncontrolled charging systems would require a corresponding scaling of the transformer capacity to ensure safety, the smart charging approach allows us to add new EVSEs without increasing the transformer capacity. To enable scalability, we can leave an open space beside each of the orignals and install a second EVSE using the same cable.  We then use the charging algorithm to ensure the capacity of this cable is not exceeded.
\iflong{
In this experiment, we assume the cable was sized for a single EVSE (32 A). However, if the scale-out was planned, a larger cable could have been installed initially.}
Thus, we can easily scale the number of EVSEs without increasing transformer or interconnection capacity. 
\iflong{
One example of this is the Caltech ACN, where 19 EVSEs were replaced in a two-for-one swap while two other EVSEs were replaced with an eight-for-one swap, scaling the system from 21 EVSEs to 54 without a change to the transformer capacity.\footnote{Note that the original 21 EVSEs had a maximum charging power of 80 A and cables were sized accordingly, while the replacement EVSEs had a maximum charging current of just 32 A.}
}

Interestingly, as the number of EVs served by the system increases, the effective cost per kWh decreases for all systems. This indicates the economics of scale, which are associated with demand charge. With more usage, it is possible to spread the demand charge over more energy delivered, decreasing the price per kWh. This decrease in demand charge is greater than the increase in energy price, which results from needing to charge users in more expensive TOU periods, leading to a net decrease in per-unit costs.

\subsection{Comparing Algorithms in Oversubscribed Infrastructure}\label{sec:oversubscribed_infrastructure}
As we have seen in Section~\ref{sec:system_planning}, smart charging algorithms can lead to significant savings in terms of both capital investment and operating costs. However, despite significant work
\iflong{in the development of algorithms for smart charging systems, }there are still relatively few algorithms proposed in the literature which can be directly applied in practice. To develop more practical algorithms, ACN-Sim provides a platform to evaluate them in as realistic a setting as possible. A key feature of \acnsim{} is its ability to simulate the unbalanced three-phase electrical infrastructure common in large charging systems. Most charging algorithms in the literature rely on constraints that implicitly assume single-phase or balanced three-phase operation. 

To see why these assumptions are insufficient for practical systems, we consider two versions of the LLF algorithm. In the first, LLF only ensures that total power draw is less than the transformer's capacity, which is sufficient for a single-phase or balanced system. In the second, LLF uses the full three-phase system model that includes individual line constraints. The results of this experiment are shown in Fig.~\ref{fig:three-phase-llf}, where we can see that only considering maximum power draw leads to significant constraint violations in line currents. However, by using an algorithm that considers the full three-phase model, we ensure these line constraints are not violated at the cost of not fully utilizing the 50 kW transformer's capacity due to unbalance.
\begin{figure}
    \centering
    \includegraphics[width=\columnwidth]{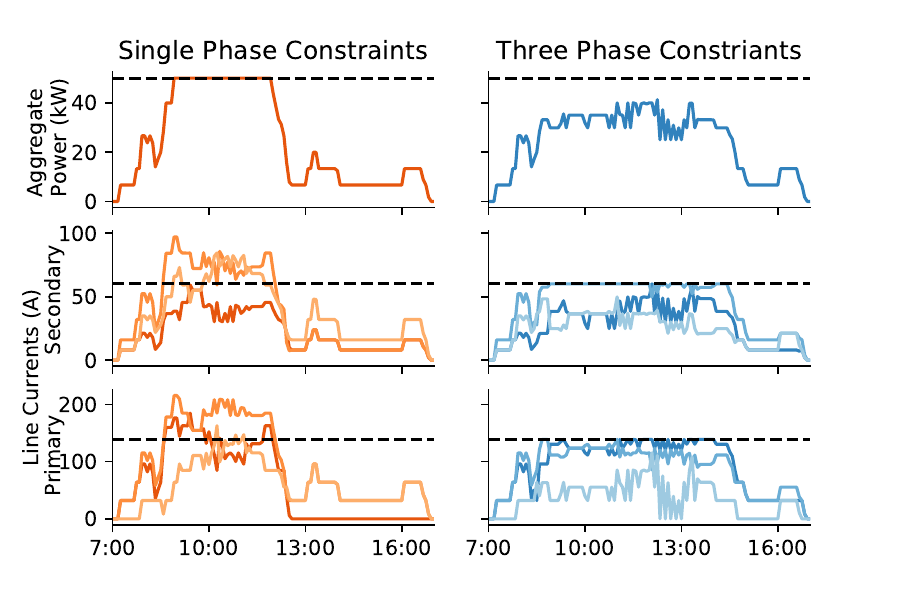}
    \vspace{-0.25in}
    \caption{Aggregate power draw and line-currents at the primary and secondary side of the transformer when running single-phase and three-phase LLF algorithms on the Caltech ACN with a 50 kW transformer capacity. Shading in the lower plots denote each phase while the black dotted line denotes the power/current limit. The experiment is based on on data from the Caltech ACN on March 5, 2019 and uses a 5 minute timestep.}
    \label{fig:three-phase-llf}
    \vspace{-0.1in}
\end{figure}

Unbalanced three-phase infrastructure can also influence our evaluation of algorithms. An important evaluation metric for EV charging algorithms is what percentage of user energy demands met when infrastructure constraints are binding. We use this metric to evaluate six algorithms over a range of possible transformer capacities based on the real charging workload of the Caltech ACN from September 2018. To demonstrate the effect of infrastructure models, we conduct this experiment with single-phase and three-phase models, as shown in Fig.~\ref{fig:infrastructure_comp}. Here we can see that in the single-phase case, EDF, LLF, and MPC all perform near optimally, exceeding the performance of Round Robin and FCFS by up to 8.6\%. However, the subplot on the right tells a different story. Here we see that the MPC algorithm can match the offline optimal performance as before, while EDF and LLF both underperform. In fact, in the highly constrained regime, Round Robin outperforms EDF and LLF despite having less information about the workload. We attribute these results to the importance of phase-balancing in three-phase systems, which has been historically under-appreciated in the manged charging literature. 

In addition to comparing algorithms, the curves in Fig.~\ref{fig:infrastructure_comp} can also inform the design of charging systems when accounting for the online algorithm used. For example, we can see that if a host wants to deliver $>$99\% of charging demand using MPC, a 70 kW transformer would be sufficient, assuming an unbalanced three-phase system. Alternatively, if an existing transformer can only support 40 kW of additional demand, a host could expect to meet approximately 85\% of demands without an upgrade. 

\begin{figure}[t!]
    \centering
    \includegraphics[width=\columnwidth]{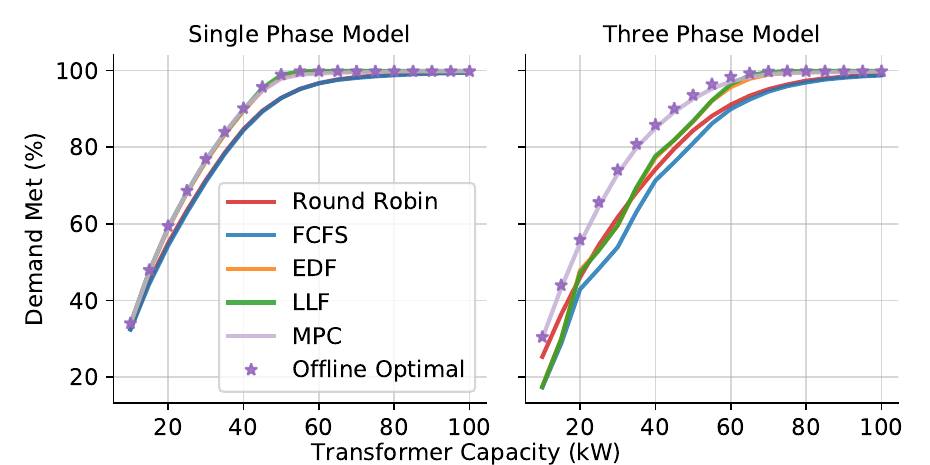}
    \vspace{-0.25in}
    \caption{Comparison of percentage of energy delivered as a function of transformer capacity for single-phase (left) and three-phase (right) systems. Stars represent the offline optimal, which is an upper bound based on perfect future information. The simulation runs from Sept. 1 through Oct. 1, 2018, with a timestep of 5 minutes. To generate events, we use \acnsim{}'s integration with \acndata{} to get real charging sessions from the Caltech ACN, assuming the ideal battery model. We also use the included Caltech ACN charging network model with ideal EVSEs, and use its optional \code{transformer\_cap} argument to limit the infrastructure capacity.
    }
    \label{fig:infrastructure_comp}
    \vspace{-0.1in}
\end{figure}

\subsection{Time Series Inspection}\label{sec:time_series}
\begin{figure}[t!]
    \centering
    \includegraphics[width=\columnwidth]{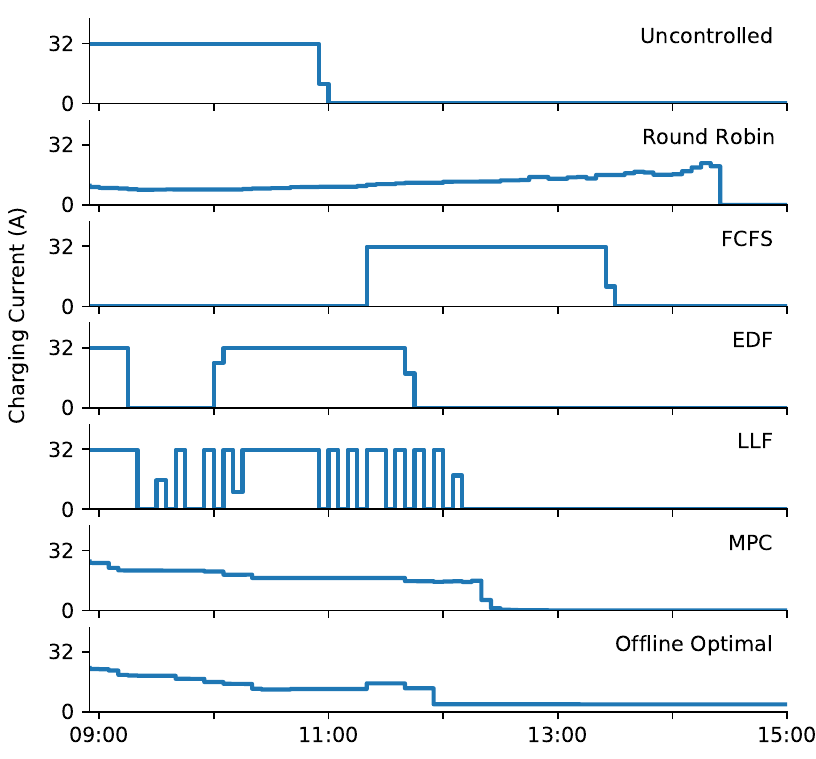}
    \vspace{-0.3in}
    \caption{Comparison of charging profiles for one EV on September 13, 2018 with a 70 kW transformer capacity.}
    \label{fig:ev_profiles}
    \vspace{-0.1in}
\end{figure}

\acnsim{} also allows us to examine the charging profile of individual EVs, as shown in Fig.~\ref{fig:ev_profiles}. Here we can see a qualitative difference between the algorithms. For example, FCFS behaves very similarly to Uncontrolled charging but is delayed as the EV must wait its turn in the queue. For EDF and LLF, charging can be interrupted when EVs with earlier deadlines arrive or as an EV's laxity evolves over time. Oscillations in the LLF plot result from an increase in laxity as the EV charges, which can decrease its standing in the queue, causing it to stop charging temporarily. Round Robin, MPC, and the offline optimal are quite different. Each EV charges steadily but at a rate below its maximum as congestion in the system necessitates sharing of charging capacity. In this case, we use an equal sharing regularizer in the objective function for MPC and Offline Optimal. See \cite{lee_large-scale_2018} for more information. With this tariff schedule, on-peak rates run from 12 - 6 pm — Offline Optimal finishes charging this user before this, while MPC goes slightly into the on-peak period.

% -------------------------------------------------------
\vspace{-0.04in}
\subsection{Grid Integration}\label{sec:grid}
\vspace{-0.04in}
\begin{figure}[t!]
    \centering
    \includegraphics[clip,width=\columnwidth]{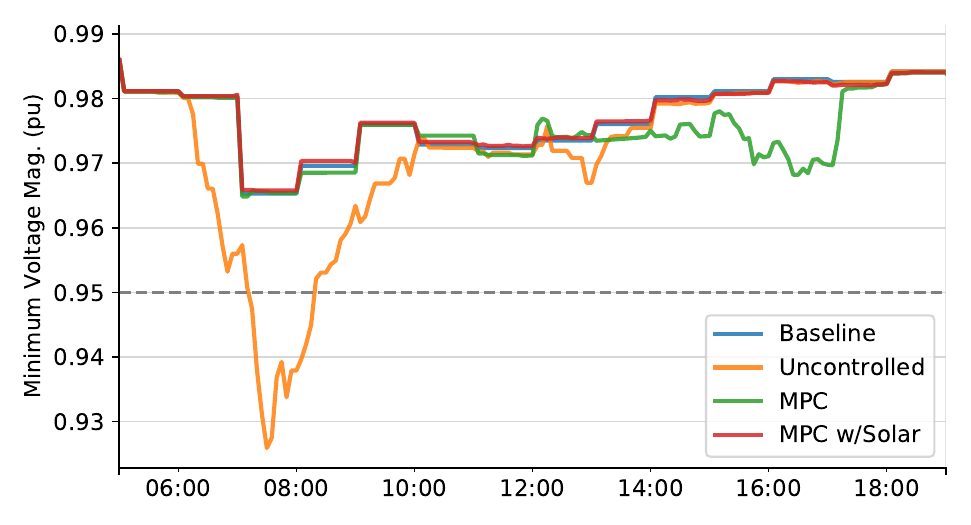}
    \vspace{-0.3in}
    \caption{Comparison of different load, generation, and charging scenarios' effects on the per unit voltage at bus 2053 in the Iowa test feeder network. The main plot shows all scenarios: from top to bottom in the legend, baseline load only, baseline load with uncontrolled EV charging, baseline load with load-flattened charging, baseline load with solar generation, and load-flattened charging. We see progressive improvements on the voltage drop, culminating in load flattening with solar almost eliminating the burden on the whole grid caused by EV charging (note that the MPC w/ Solar and Baseline curves are nearly co-located).}
    \label{fig:allvm}
    \vspace{-0.1in}
\end{figure}
The output load profiles generated by \acnsim{} simulations can be used as loads on nodes of a distribution feeder in a grid simulator; in this case study, we use OpenDSS \cite{epri_opendss}. This functionality can be used to evaluate different algorithms' effects on a larger grid.
For this case study, we add EV charging to one node of a 240-node test distribution feeder with a voltage regulator located in Iowa and use actual smart-meter data from the system in 2017\cite{bu_2019}.

We add an EV charging facility as an unbalanced three-phase load to node 2053, which has a transformer capacity of 225 kVA. We use the JPL charging network with workload data from Sept. 5, 2019. Background load data is from Sept. 5, 2017 (both days were weekdays). We consider four cases, a baseline with no EV charging, uncontrolled charging, MPC with a load flattening objective, and MPC with load flattening and onsite solar. The results of these experiments are shown in Fig.~\ref{fig:allvm}. Uncontrolled charging results in an unacceptable minimum voltage of under 0.93 p.u. and overloads the transformer at bus 2053. This indicates that the grid as designed could not support uncontrolled charging at this scale at bus 2053. 

To prevent voltage issues, we can schedule charging during periods of low background load by using the MPC framework in Section~\ref{sec:included_alg} with a load flattening objective term. We provide the actual building load as an input to the algorithm and ensure that the total load is constrained to be below the transformer's capacity. From Fig.~\ref{fig:allvm}, we can see that this improves the minimum system voltage to 0.965 p.u., which matches the system wide minimum from the baseline case. 

Since many EV charging systems are co-located with solar PV, we now consider adding a 270 kW DC (225 kW AC) PV array at node 2053. The solar data was generated from NREL's SAM tool for Des Moines, Iowa, in a typical meteorological year (TMY) for Sept. 5. We use the same MPC algorithm but now set the background load to the net load after subtracting solar. We see in Fig.~\ref{fig:allvm} this roughly recovers the same grid-wide minimal voltage as before we added an ACN, indicating that smart charging and solar PV could enable widespread adoption of EV charging without adverse grid impacts. The MPC objective is constructed to fulfill all EVs' energy demand in both the baseline and solar cases. Both of these case studies assume perfect knowledge of background load and generation, as forecasting methods are beyond the scope of this study. However, no knowledge of future EV arrivals is used.

% -------------------------------------------------------
\vspace{-0.025in}
\section{Conclusions}\label{sec:conclusion}
\vspace{-0.025in}
In this work, we present \acnsim{}, a data-driven simulator designed to aid the development of practical online scheduling algorithms for EV charging. This tool significantly reduces the software engineering burden on researchers and exposes them to practical issues present in real charging systems. \acnsim{} also makes it easier for researchers to share their experiment code, improving transparency and code reuse in the community. Finally, \acnsim{} integrates with the Adaptive Charging Network Research Portal, a larger suite of tools that includes a database of real charging sessions and a framework for field testing algorithms. \acnsim{} will continue to grow to meet the needs of the community, including new models of systems components and charging networks.

\iflong{
We also plan to develop a suite of test cases (include networks and scenarios) which can be used as a standard evaluation set for algorithms.}

\vspace{-0.025in}
\bibliographystyle{ieeetr}
\scriptsize {
\bibliography{my_library}
}
\end{document}